# A reconstruction algorithm for electrical capacitance tomography via total variation and $l_0$-norm regularizations using experimental data

**Jiaoxuan Chen[1,2], Maomao Zhang[1] and Yi Li[1,3]**

[1]Graduate School at Shenzhen, Tsinghua University, Shenzhen 518055, China

[2] Department of Automation, Tsinghua University, Beijing 100084, China

[3]Author to whom any correspondence should be addressed.

E-mail: liyi@sz.tsinghua.edu.cn

## Abstract

Electrical capacitance tomography (ECT) has been investigated in many fields due to its advantages of being non-invasive and low cost. Sparse algorithms with $l_1$-norm regularization are used to reduce the smoothing effect and obtain sharp images, such as total variation (TV) regularization. This paper proposed for the first time to solve the ECT inverse problem using an $l_0$-norm regularization algorithm, namely the doubly extrapolated proximal iterative hard thresholding (DEPIHT) algorithm. The accelerated alternating direction method of multipliers (AADMM) algorithm, based on the TV regularization, has been selected to acquire the first point for the DEPIHT algorithm. Experimental tests were carried out to validate the feasibility of the AADMM-DEPIHT algorithm, which is compared with the Landweber iteration (LI) and AADMM algorithms. The results show the AADMM-DEPIHT algorithm has an improvement on the quality of images and also indicates that the DEPIHT algorithm can be a suitable candidate for ECT in post-process.

*Keywords: electrical capacitance tomography, $l_0$-norm regularization, total variation*

# 1. Introduction

Multiphase flow imaging occurs in a variety of industrial processes and plants including petroleum, chemical and power industries. Electrical tomography (ET), such as electrical capacitance tomography (ECT) and electrical resistance tomography (ERT), is considered a highly promising technique. Thus, ET has witnessed widespread application in the past [1][2][3]. Besides being non-radioactive and non-invasive, ECT provides the advantages of being low cost and high process speed. Currently, ECT is a powerful process-imaging technique to reconstruct the permittivity distributions based on the measured capacitances between each pair of electrodes in an ECT sensor. However, ECT has the major drawback of offering low resolution images due to the inherence of ill-posedness, ill-conditioning and non-linearity.

Many algorithms have been proposed to solve ECT inverse problem [4] and the most widely-used one-step algorithm and iterative algorithm is linear back projection (LBP) [5] and the Landweber iteration (LI) [6], respectively. The inverse problem in ECT is severely ill-posed, therefore the regularization is needed. Tikhonov regularization is a typical method to solve the ECT inverse problem based on the $l_2$-norm regularization [7]. However, this method leads to the reconstructed images smoothed excessively. Recently, a sparse reconstruction with $l_1$-norm regularization is used to reduce the smoothing effect and to obtain sharp images, such as total variation (TV) regularization. In the past few years, the TV method for ECT imaging has received considerable attention: Soleimani and Lionheart [8] explored a regularized Gauss-Newton scheme and found that the TV regularization showed distinctive advantage in obtaining sharp images; Hosani *et al* [9] presented different algorithms to reconstruct the high contrast objects and found that the TV method showed better results compared with the Tikhonov regularization method; Ye *et al* [10] designed an unconventional basis for ECT, which is based on an extended sensitivity matrix; Chen *et al* [11] introduced two numerical methods to solve the imaging problems in ECT based on Rudin–Osher–Fatemi (ROF) model with TV regularization.

Chen *et al* proposed an iterative algorithm for ECT based on TV regularization, namely accelerated alternating direction method of multipliers (AADMM) [11]. They concluded that the AADMM algorithm could identify the object from its background efficiently and make the boundary of the object clear in several cases. However, they also pointed out that some artifacts in the images reconstructed by the AADMM could not be removed. The $l_1$-norm based approaches are capable of obtaining a sparse solution by using a soft thresholding operator. On the other hand, these approaches yield loss of contrast and eroded signal peaks [12]. The $l_0$-norm regularization has its advantages over $l_1$-norm regularization in many applications[13][14][15].However, the feasibility of $l_0$-norm based approach for improving image quality has not been assessed for ECT.

Although the AADMM algorithm could distinguish the edge of the object effectively, the reconstructed permittivity over the region of the object is not homogenous. The existing post-process method to deal with it is binarization of the images with setting thresholds. This method is rough and sometimes may make a damage to the original images. In addition, the detailed values of those area sometimes cannot be attained, *i.e.* the thresholds cannot be determined. The main motivation of this paper is to improve the quality of the images reconstructed by the AADMM algorithm. In this paper, a combined algorithm for ECT via total variation and $l_0$-norm regularizations is proposed. The algorithm consists of two steps: the first step is to use the AADMM algorithm to obtain the initial solution; the second is to use the DEPIHT algorithm to reduce the artifacts in the images and then enhance the intensity over the blurred area.

This paper is organized as follows: in section 2, inspired by the previous research [11][16], a combined algorithm for ECT is introduced; Section 3 describes the experimental setup. Results and discussion of experimental data are provided in section 4 to validate the feasibility of the proposed algorithm, and section 5 concludes the paper.

## 2. Princilple of algorithm

Bao *et al* presented an $l_0$-norm based algorithm, namely extrapolated proximal iterative hard thresholding (EPIHT) [16]. Inspired by this work, we propose the doubly extrapolated proximal iterative hard thresholding (DEPIHT) for ECT. Since the DEPIHT for solving $l_0$-norm regularization problem can merely guarantee local convergence, the initial point for DEPIHT is needed. The TV regularization is able to gain a good shape recovery in ECT reconstruction. Thus, the AADMM algorithm is used to acquire the initial point for DEPIHT. The process of the AADMM-DEPIHT algorithm is shown in figure 1.

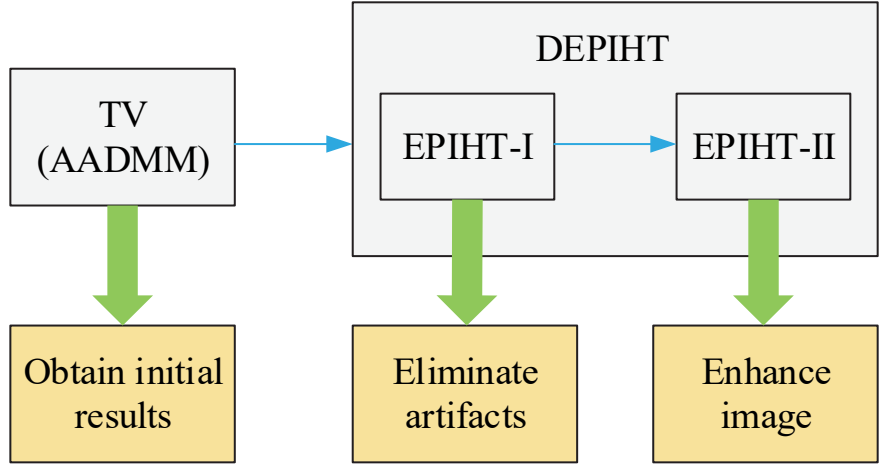

Figure 1. Process diagram of the AADMM-DEPIHT algorithm

In ECT, the mathematical model between the capacitance and permittivity distributions can be represented as [17][18][19]

$$\lambda = Sg \tag{1}$$

where $\lambda$ is a normalized capacitance, $S$ is a normalized matrix known as the sensitivity map, and $g$ is the normalized permittivity.

The general equation of the AADMM algorithm can be transformed from the equation (1) using an optimization perspective.

$$\min_{g} \frac{\mu}{2}\|Sg - \lambda\|_2^2 + \frac{\varepsilon}{2}\|\nabla g\|_2^2 + \|\nabla g\|_1 \tag{2}$$

where the first term is the fidelity term with parameter $\mu$, the third term is a TV term, $\varepsilon$ is a smoothing parameter, $\nabla$ is a gradient operator. A full description of the AADMM algorithm has been published in [11], therefore only a brief summary of this algorithm is given here.

The DEPIHT algorithm consists of two steps: the first step is the EPIHT-I algorithm, the second is the EPIHT-II algorithm. Firstly, an optimization case for ECT based on $l_0$-norm regularization is given as below,

$$\min_{g} \frac{1}{2}\|Sg - \lambda\|_2^2 + \frac{1}{2}\|g\|_2^2 + r\|g\|_0 \tag{3}$$

where $r$ is a non-negative sparsity-promoting weight parameter.

Then, define two functions $H(g)$ and $G(g)$,

$$\begin{cases} H(g) = G(g) + r\|g\|_0 \\ G(g) = \frac{1}{2}\|Sg - \lambda\|_2^2 + \frac{1}{2}\|g\|_2^2 \end{cases} \tag{4}$$

And the surrogate function $R_q(x,y)$ of $H(g)$ is set up as

$$R_q(x, y) = r\|g\|_0 + G(y) + \langle \nabla G(y), x - y \rangle + \frac{q}{2}\|x - y\|_2^2 \tag{5}$$

where $q$ is a non-negative parameter. The AADMM-DEPIHT algorithm is shown in algorithm1 explicitly.

| |
|---|
| Algorithm1. AADMM-DEPIHT |
| AADMM step: |
|    1. Obtain an initial solution by using the AADMM algorithm, *e.g.* $g_{tv}$ |
| EPIHT-I step: |
|    2. Inputs: sensitivity matrix $S$, capacitance measurements $\lambda$, a parameter used in the extrapolation step $w$, the number of iterations $k_{max}$ and two parameters $r$, $q$. |
|    3. Initialize $g_{-1} = g_0 = g_{tv}$ and $k=0$. |
|    4. While $k < k_{max}$<br><br>$$y_{k+1} = g_k + w(g_k - g_{k-1})$$<br><br>    if $H(y_{k+1}) > H(g_k)$<br><br>$$y_{k+1} = g_k$$<br><br>    end if<br><br>$$g_{k+1} \in \arg\min_{g} R_q(g, y_{k+1}) \quad\quad (6)$$<br><br>$$k = k+1$$<br><br>   end while |
|    5. Let $g_{tv} = g_{k_{max}}$ |
| EPIHT-II step: |
|    6.<br><br>    *repeat the steps from 2 to 4 except for the equation (6):*<br><br>$$g_{k+1} \in \arg\min_{g} R'_q(g, y_{k+1}) \quad\quad (7)$$<br><br>    (the relationship between $R_q$ and $R'_q$ will be concerned in the following.) |
|    7. Output: $g_{k_{max}}$. |

The equation (6) is given by

$$g_{k+1} \in \psi_{\sqrt{\frac{2r}{q}}}\left(y_{k+1} - \frac{1}{q}\nabla G(y_{k+1})\right) \tag{8}$$

where $\psi_a(\cdot)$ denotes the hard thresholding operator, which is defined as below.

$$[\psi_a(b)]_i = \begin{cases} [b]_i, & |[b]_i| > [a]_i \\ 0, & \text{else} \end{cases} \tag{9}$$

where $[\cdot]_i$ denotes the $i$th component of a vector. The equation (7) is given by

$$g_{k+1} \in \upsilon_{\sqrt{\frac{2r}{q}}}\left(y_{k+1} - \frac{1}{q}\nabla G(y_{k+1})\right) \tag{10}$$

where $\upsilon_a(\cdot)$ is analog with the hard thresholding operator, which is expressed as

$$[\upsilon_a(b)]_i = [b]_i \text{ if } |[b]_i| > [a]_i \tag{11}$$

In fact, the relationship between $R_q$ and $R'_q$ has little difference except for the meanings of the thresholding operator. However, this leads to a significantly different effect on the ECT reconstruction.

## 3. Experimental setup

Figure 2 illustrates a typical ECT system, which comprises mainly of three subsystems: a typical ECT sensor with eight electrodes, a data acquisition device and a computer. In the test, the diameter of the ECT sensor was 76 mm and the angular span of each electrode was 30°. The data acquisition speed of the data acquisition system was about 350 Hz, *i.e.* the data acquisition system can acquire about 350 sets of capacitance data of

28 electrodes pairs per second. The signal-to-noise ratio (SNR) of capacitance data for each of 28 electrodes pairs ranged from 30 dB to 40 dB. In order to avoid the system errors and noises, the average of thousands of frames was employed as the capacitance data. The imaging was completed using MATLAB R2015b on the computer with an Intel Core i5-6400 2.7 GHz CPU and 4 GB of RAM.

Four distributions are set for the test: cross-shaped, 'V'-shaped, two rectangular-shaped and three circular-shaped, as shown in Figure 3. Air and dry sand were used as the low and high permittivity materials (relative permittivity 1 and 4 respectively) to calibrate the system. Paper-made containers in different shapes are filled with dry sand, which represent for each tested distribution. In addition, a mesh with 2304 (48×48) square elements is used to generate the sensitivity matrix $S$ with an air in the measuring space.

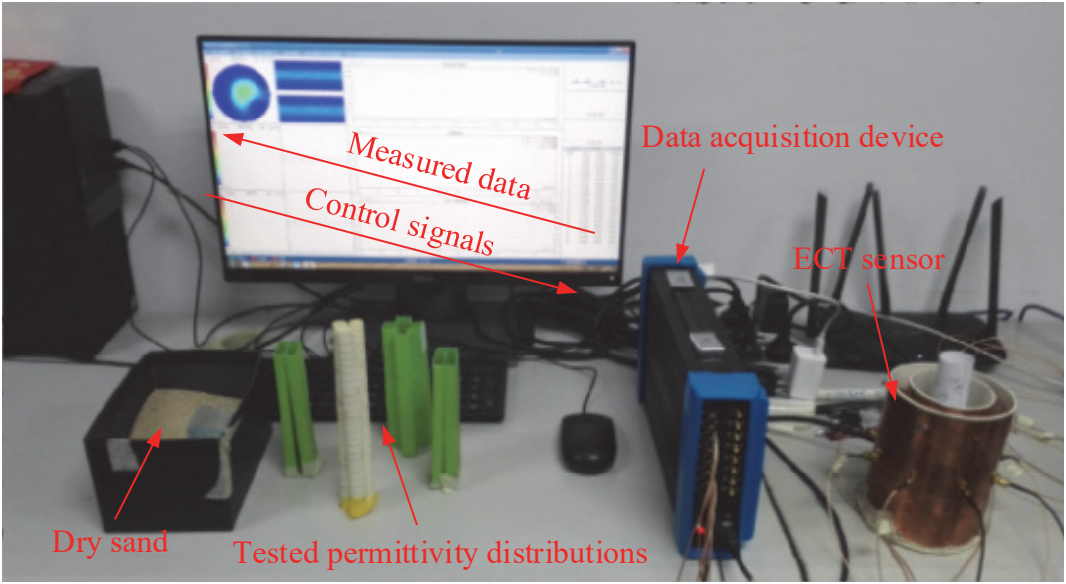

Figure 2. ECT system and materials used in the experimental test.

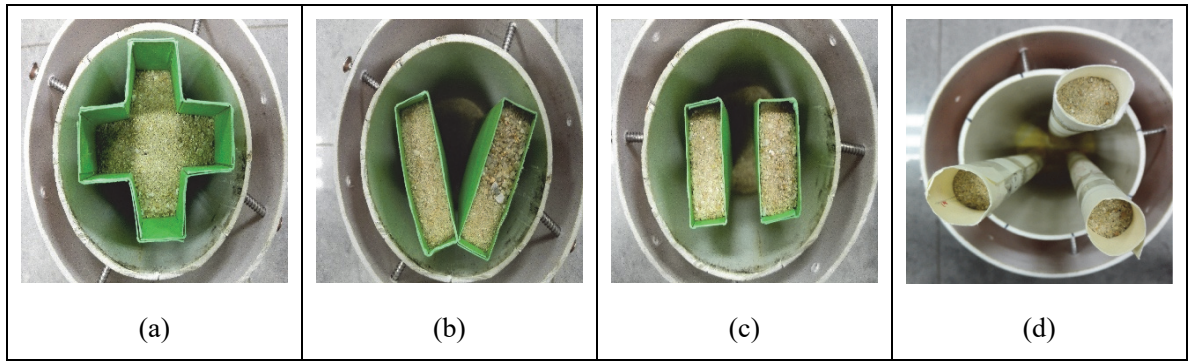

Figure 3. Real permittivity distributions used in the experimental test : (a) cross-shaped, (b) 'V'-shaped, (c) two rectangular-shaped and (d) three circular-shaped

Parameter selection is essential to the quality of reconstructed images. Some initial parameters of the three algorithms are stated here. The relaxation factor and the number of iteration for LI are chosen as 0.008 and 3000 respectively. This can ensure the LI algorithm converge at a good point. The parameter selection for the AADMM algorithm can be suggested in [11] and will not be discussed in this paper. In the AADMM-DEPIHT algorithm, the parameter selection for the EPIHT-I and EPIHT-II algorithms is different. In the EPIHT-I algorithm, the parameter $r$ is set to 0.01 while the parameter $w$ is chosen to be 1000. In the EPIHT-II algorithm, the parameter $r$ is set to 1 while the parameter $w$ is chosen to be 1. Moreover, the parameter $q$ used in the EPIHT-I and EPIHT-II algorithms varies for the different distributions, which can be regarded as a controlling parameter.

## 4. Results and discussion

Figure 4 provides a 2D images reconstructed from the experimental data. It can be found that the AADMM algorithm can identify the objects from the background efficiently, especially in the test 2. In the test 4, although the LI and AADMM algorithms can both distinguish the three circular objects from the background, there are some distinct artifacts in the result of the LI algorithm.

The existing binary process for ECT is to use a threshold method, *e.g.* threshold operator (TO), which can be expressed as below.

$$TO(thr) = \begin{cases} 0 & g_i < thr \\ 1 & g_i > thr \end{cases} \qquad (12)$$

where *thr* denotes the value of the threshold.

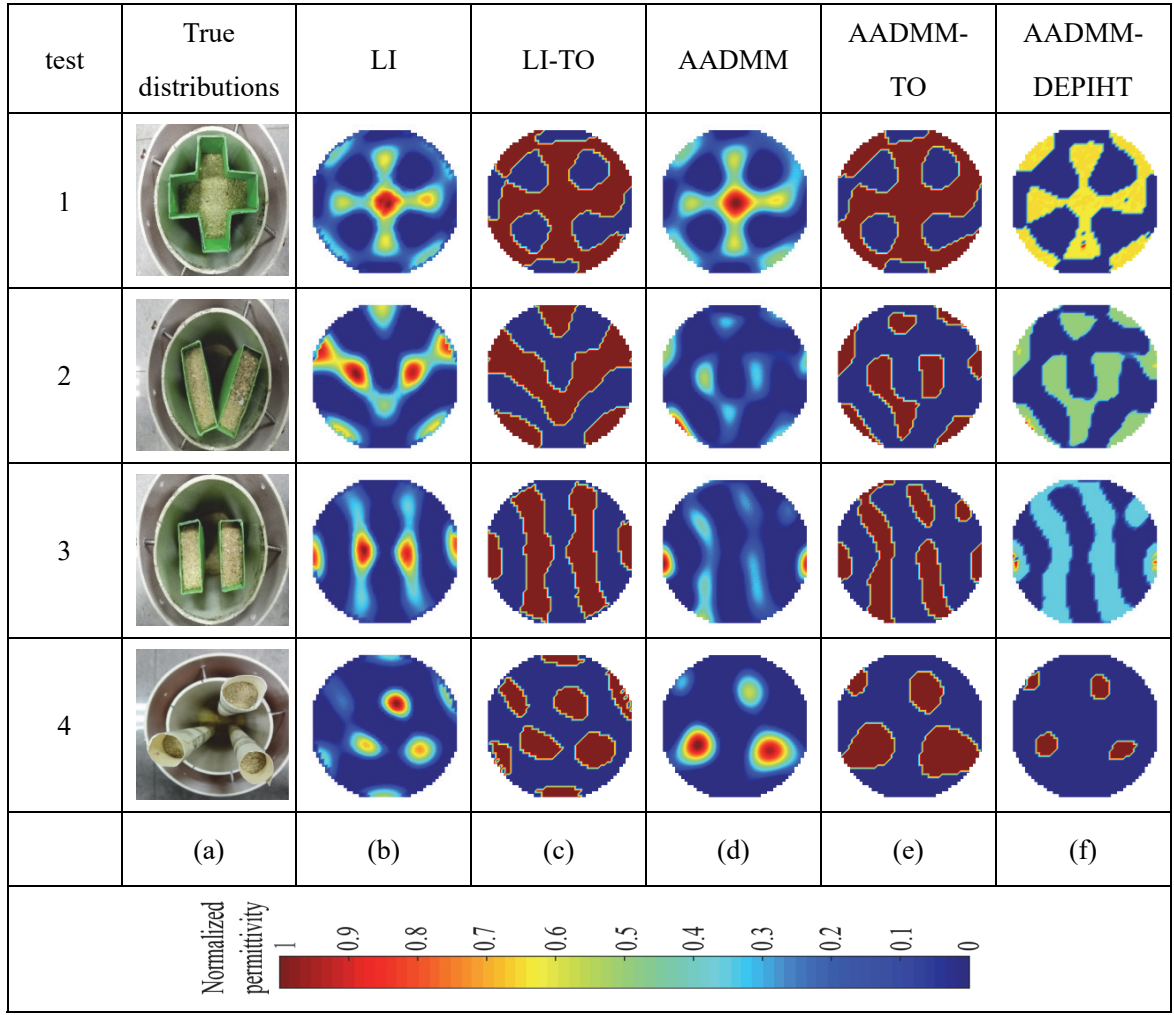

Figure 4. 2D images reconstructed from the experimental data : (a) the true distributions, (b) the images reconstructed from the LI algorithm, (c) the images reconstructed from the LI algorithm with a threshold, (d) the images reconstructed from the AADMM algorithm, (e) the images reconstructed from the AADMM algorithm with a threshold and (f) the images reconstructed from the AADMM-DEPIHT algorithm.

As is shown in figure 4, the cases (c) and (e) show the results of the LI and AADMM algorithms with the operator of TO, respectively. The parameter *thr* used in the TO is chosen to be 0.1 for all distributions. It is worth noting that the original images are normalized between 0 and 1. Hence, the only prior knowledge is the range of the reconstructed permittivity while using the threshold method. Figure 4 provides that although the threshold method can make the objects in the images more clear, this method sometimes makes a damage to the original objects, showing the threshold method is rough. Naturally, the selection for the value of the threshold has a deep influence on the final results, which shows this method is unpredicted.

Regarding to the results of the AADMM-DEPIHT algorithm, the images are very vivid. To quantitatively evaluate the fluctuation of each imaging result, the standard deviation (SD) in equation (13) is calculated. From the figure 5, it can be found that the SD of the AADMM-DEPIHT algorithm is far less than that of the LI and AADMM algorithms, indicating the superiority of this algorithm.

$$\sigma = \sqrt{\frac{1}{M}\sum_{i=1}^{M}(g_i - g_0)^2}$$
$$g_0 = \frac{1}{M}\sum_{i=1}^{M} g_i$$
(13)

where $\sigma$ is the standard deviation, $M$ is the total number of pixels in an image, $g_i$ is the reconstructed permittivity value and $g_0$ is the mean value of reconstructed image.

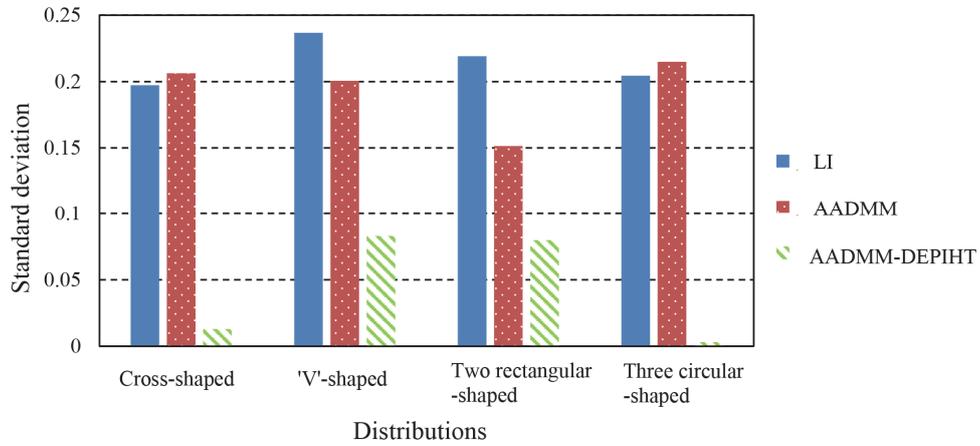

Figure 5. The standard deviation of the results reconstructed by the LI, AADMM and AADMM-DEPIHT algorithms.

Figure 6 provides a 3D images corresponding to the figure 4. From the figure 6, it can be found that a main advantage of the DEPIHT algorithm is to enhance the images despite the artifacts. This algorithm intends to lift the non-zero permittivity to the similar height, *i.e.*, it turns the "hills" into "pillars". Thus, it is of importance to reduce the artifacts in the images during the process of the DEPIHT algorithm. The EPIHT-I algorithm is used to reduce the artifacts. It seems this step plays a substantial role in the test 1 and 4 while has a little effect in the test 2 and 3. However, compared to the AADMM algorithm, the AADMM-DEPIHT algorithm has an improvement on the quality of images to some degree. Perhaps, removing the artifacts in the images could be investigated in the future, enabling better results in the AADMM-DEPIHT algorithm. In fact, the DEPIHT can be regarded as a post-process method for ECT. For instance, the LI algorithm is used to acquire the first point and then the method becomes the LI-DEPIHT algorithm.

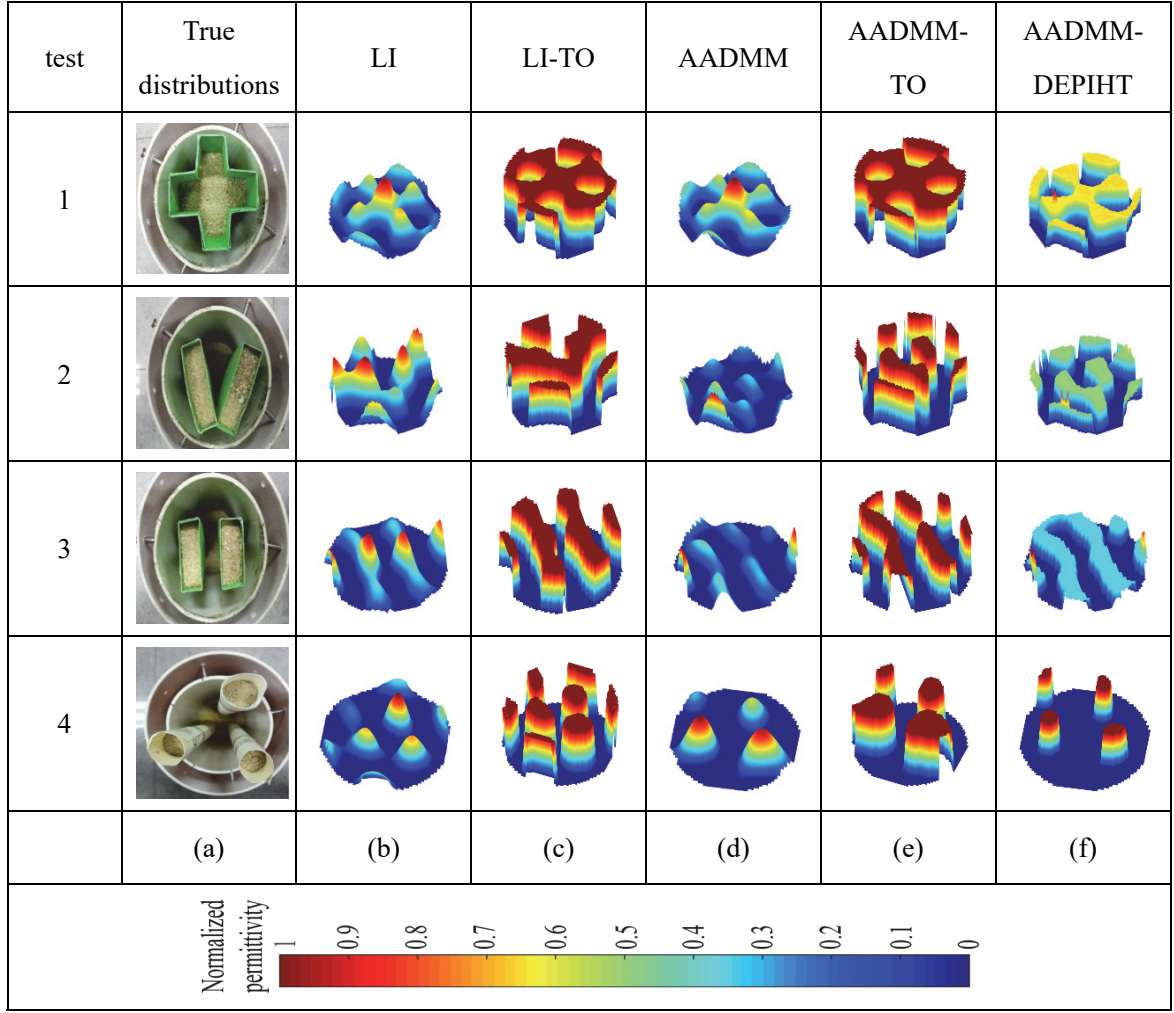

Figure 6. 3D images reconstructed from the experimental data : (a) the true distributions, (b) the images reconstructed from the LI algorithm, (c) the images reconstructed from the LI algorithm with a threshold, (d) the images reconstructed from the AADMM algorithm, (e) the images reconstructed from the AADMM algorithm with a threshold and (f) the images reconstructed from the AADMM-DEPIHT algorithm.

Table 1: Elapsed time (in seconds)

| Distributions | LI | AADMM | AADMM-DEPIHT |
|---|---|---|---|
| Cross-shaped | 0.47 | 1.22 | 1.37 |
| 'V'-shaped | 0.45 | 2.41 | 2.53 |
| Two rectangular-shaped | 0.46 | 2.51 | 2.63 |
| Three circular-shaped | 0.47 | 0.71 | 0.85 |

Table 1 shows the elapsed time of the three algorithms. Since the mesh used in this study is 2304 (48×48) square elements, the imaging time of the LI and AADMM algorithms decreases a lot compared in [11]. As shown in table 1, the elapsed time of the DEPIHT algorithm is very short, which is benefited from the hard thresholding operator. This makes the DEPIHT algorithm a suitable candidate for ECT in post-process.

## 5. Conclusion

A combined algorithm for ECT via total variation and $l_0$-norm regularizations, namely the AADMM-DEPIHT algorithm, is presented and validated using experimental data. The results show the AADMM-DEPIHT algorithm does have an improvement on the quality of images compared to the AADMM algorithm, *e.g.* remove some artifacts in several cases. Furthermore, the DEPIHT algorithm, based on $l_0$-norm regularization, has a very suitable application in binarization, *e.g.* a post-process for ECT. However, the DEPIHT algorithm cannot identify the object from background while enhancing the image. Thus, the step of reducing major artifacts is necessary. It is anticipated that $l_0$-norm regularization methods such as the DEPIHT algorithm, can be combined with other global convergence algorithms, enabling better reconstruction in ECT.

## Acknowledgements

The authors would like to thank the National Natural Science Foundation of China (No.61571252) for supporting this work.